# Observers' Pupillary Responses in Recognising Real and Posed Smiles: A Preliminary Study

## Research-in-progress

### Ruiqi Chen
Research School of Computer Science
The Australian National University
Canberra, Australia
Email: u6660094@anu.edu.au

### Atiqul Islam
Research School of Computer Science
The Australian National University
Canberra, Australia
Email: atiqul.islam@anu.edu.au

### Tom Gedeon
Research School of Computer Science
The Australian National University
Canberra, Australia
Email: tom@cs.anu.edu.au

### Md Zakir Hossain
Research School of Computer Science
The Australian National University
Canberra, Australia
Email: zakir.hossain@anu.edu.au

## Abstract

Pupillary responses (PR) change differently for different types of stimuli. This study aims to check whether observers' PR can recognise real and posed smiles from a set of smile images and videos. We showed the smile images and smile videos stimuli to observers, and recorded their pupillary responses considering four different situations, namely paired videos, paired images, single videos, and single images. When the same smiler was viewed by observers in both real and posed smile forms, we refer them as "paired"; otherwise we use the term "single". The primary analysis on pupil data revealed that the differences of pupillary response between real and posed smiles are more significant in case of paired videos compared to others. This result is found from timeline analysis, KS-test, and ANOVA test. Overall, our model can recognise real and posed smiles from observers' pupillary responses instead of smilers' responses. Our research will be applicable in affective computing and computer-human interaction for measuring emotional authenticity.

**Keywords** Physiological Signals, Pupillary Responses, Real Smiles, Posed Smiles, KS test.

# 1  Introduction

The dilation and constriction of human pupils are not merely the indication of response to luminance but also are an importance indication of psychophysiology. The aperture of the pupil conveys interesting information regarding human internal cognitive states by receiving input from sympathetic and/or parasympathetic nervous systems. Pupil dilation and constriction caused by sympathetic and parasympathetic excitation, respectively or parasympathetic and sympathetic inhibition, respectively (Cacioppo et al. 2007). Thus, the relation between pupillary responses and cognitive processing has been a subject of interest since the early 60s (Hess and Polt 1964). However, the abundant literature on the relationship of pupillary responses and cognitive processing in the last decade is an indication of the current popularity of this particular physiological signal (Ariel and Castel 2014; Palinco et al. 2010). The majority of recent research measured pupillary responses through video-based eye tracking and pupillometric feature extraction and analysis (Huh et al. 2019; Van et al. 2019; Chen et al. 2011; Xu et al. 2011).

This study takes an initial step to analyse the problem of recognising real and posed smiles based on observers' pupillary response data under certain conditions. The smile is one of the simplest human facial expressions, and yet would appear to be one of the most complex displays of expression. The literature shows some studies have reported successful computational technique to recognize smiles with good accuracy. However, in computer vision, a smile is depicted by as little as lip corner pull and cheek raise. People smile for many reasons beyond happiness, such as for embarrassment, anxiety, depression, deep sadness and many more. Moreover, it can be said that the smile is a plausible indication of original emotion in the context of social display. In this context, recognising a genuine smile from 'fake' ones can help us to know many important aspects of human emotion. Several studies have attempted to distinguish between real and posed smiles (Dibeklioğlu et al. 2012; Hoque and Picard 2011; Yang et al. 2020). Recognising the smile nature or veracity has many potential applications in customer service, cognitive behaviour modelling, information systems, and many more (Chen et al. 2017; Yang et al. 2020).

Nowadays, there is a great deal of application of user behaviour intention, psychology, affect detection, and many more on Information Systems (Sun and Zhang 2006; Willis and Jones 2012). Our current study focuses on determining how an observer's pupillary responses reflect their responses to smile videos and images to distinguish real and posed smiles. It influences objective understanding of how observers' pupillary responses react to real and posed smiles that they witness in everyday environments. Pupillary responses could also be useful in many potential applications in information systems that are beyond our study. For example, task evoked pupillary responses have been used by clinicians and other relevant researchers to assess patient's early risk for mild cognitive impairment (MCI) and Alzheimer's disease (AD) (Granholm et al. 2017). Researchers have used pupillary responses for cognitive processing in visualization of different graph and data (Hossain et al. 2018). It may also be possible to detect and regulate individuals' emotion from pupillary responses (Kinner et al. 2017). We use participants and record their pupillary responses while observing other people's smiles on a screen. This has a wide range of future potential applicability to Information System applications, such as affect and personality detection (Berkovsky et al. 2019).

Previously, smile videos were analysed to distinguish between real and posed smile (Valstar et al. 2007). Virtual avatars of smiling faces have been studied in (Ochs et al. 2012). Dynamic and morphological characteristics of smiles are studied in (Ambadar et al. 2009). The authors have analysed the facial images and facial action coding systems to measure the characteristics of smiles. While those studies are quite influential in the field of smile detection, none of them attempted to understand the smiles from a human factors perspective. In our study we try to analyse an observer's pupillary responses while they watch (a pair of videos and/or images of) a smiling face. We put forward a hypothesis that observers' pupillary response can be used to distinguish between real and posed smiles instead of analyzing smile faces themselves. Thus, the research will help us to develop an app for monitoring caller smiles by the recipient to enhance emotional communication as we developed a deep network, called RealSmileNet, for smile recognition (Yang et al. 2020) and a correlation model, called CCA network, for selecting features from physiological/biomedical data (Hossain et al. 2016).

## 2  Experimental Methodology

The main focus of our study is to analyse the pupillary responses produced by cognitive processing as evoked by observing real and posed smiles.

### 2.1  Smile Image and Video Data

The smile data were collected from a recent database called the UvA-NEMO dataset (Dibeklioglu and Salah 2015). The UvA-NEMO (University of Amsterdam-NEMO) is a well-known large-scale smile database that contains 1,240 smile videos from 400 participants (185 female, 215 male participants) from an age range of 8 to 76 years. There are 597 real smile and 643 posed smile videos. All the videos are recorded under controlled illumination conditions. A total of 60 videos were randomly selected for this experiment. We created four categories from these 60 videos, namely PV (paired videos), PI (paired images), SV (single video), and SI (single image). In a random selection 60 videos created enough variation for our observers within the scope of the experiment for this current study. When the same person was viewed by observers in both real and posed smile videos / images, we report this as "paired smiles", otherwise we use the term "single smile". Here, image means a single frame and video means a sequence of frames. The selected videos / images were processed using oval masks to keep the face portion only, and presented to the observer in an order balanced way to avoid any order effects (Hossain and Gedeon 2017). We kept the length of each video the same as the source video in the dataset. The length of the videos usually spans from a minimum of 2s to a maximum of 8s, being the natural length of that smile. In the case of images, we chose one of the middle frames from each video and showed the frame for five seconds. Overall, we considered 20 videos for PV, 10 videos for SV, 20 images for PI, and 10 images for SI respectively.

### 2.2  Participants / Observers

We have recruited 25 volunteers as participants (also called observers as they watch or observe the smile images and/or videos) in the experiment who were University students. Amongst the participants, we had 13 male and 12 female observers with age range of 24.6±2.54 (average ± standard deviation). From the literature we can see that even less than 25 participants provide reliable outcome for a similar kind of research (Hossain and Gedeon 2017). All the participants had normal or corrected to normal vision and provided written consent prior to their participation. We have asked our participants about their vision because if they do not wear their prescribed glasses or contact lens, they might have difficulty seeing the video stimuli properly which will influence the result in terms of cognitive arousal. Approval from our University's Human Research Ethics Committee was received before the experiment.

### 2.3  Experimental Procedure

For this study we have collected participants'/observers' pupillary responses (PR) data while performing the experimental activities. The Eye Tribe[1] remote eye-tracker was used to record the PR signals at a sampling rate of 60 Hz. Observers were briefed about the experiment after arrival at the laboratory, and asked to sign the consent form. They were seated on a static chair, facing a 15.6-inch ASUS laptop in a sound-attenuated, dimly lit, closed room. The Eye Tribe remote eye tracker was used to record their eye data. Their chairs were moved forward or backwards to adjust the distance between the chair and laptop. Observers were asked to limit their body movements in order to reduce undesired artefacts in the signals. The smile videos or images were presented to the observers in a randomised fashion considering each experimental condition separately (i.e. SI, SV, PI, and PV). At the end of each (or pair) video or image, the observer was asked to choose whether the smile was real or posed by answering the question on the laptop into a web interface. The total duration of the experiment was around 45 minutes. After completing the experiment, the sensors were removed, and the observers were thanked for their participation. We have stored the experimental data for future processing. Filtering, interpolation and normalization were applied as a pre-processing on each signal. After that we have applied statistical analysis on the processed signal.

### 2.4  Data Processing

We have removed the participant data that contains any erroneous signal excluding eye blinks, which is a natural phenomenon. Therefore, there were 24 observers' data to process and analyse. As pupillary responses are affected by eye blinks, we needed to reconstruct the missing values caused by the eye blinks. There is evidence that the average human eye blink period ranges from 0.1s to 0.4s (Schiffman 1990). So, if the missing data exceeds about 1.0s, they are not considered for further analysis. Data interpolation are implemented only for shorter consecutive missing data which are mainly caused due to eye blinks. Cubic spline interpolation was applied in this case to reconstruct the missing values (Sebastiaan 2013). The reconstructed data was smoothed using Hampel filters to remove any further artefacts (Liu et al. 2004). Then, a popular normalisation method, called baseline normalization, was applied for reducing age effects on the data (Mathot et al. 2018). The baseline normalisation is mathematically expressed as follows (equation 1) where median pupil size during 100 samples prior to a test are treated as the baseline.

---

[1] https://theeyetribe.com

$$Normalisation = \frac{Pupillary\ Response - Baseline}{Baseline} \quad (1)$$

The normalized data was further smoothed using moving window average of length 11. As videos' lengths varied from 2s to 8s, we resampled them to make each of them 5s long (Akima 1970). Finally, we moved the intercept of the y-axis to the same point for comparison purposes.

To investigate the differences between real and posed smiles for all the data, the KS test (Blair and Karniski 1993) was performed to report significance level and an analysis of variance (ANOVA) (Dunn and Smyth 2018) was performed in individual conditions with the assumption that the real smile signal and posed smile signal follow the same distribution, and *p*-value less than 0.05 indicates that they do not follow the same distribution. The KS test is one of the most general and useful methods for comparing two samples distributions. The data processing steps are shown in Figure 1.

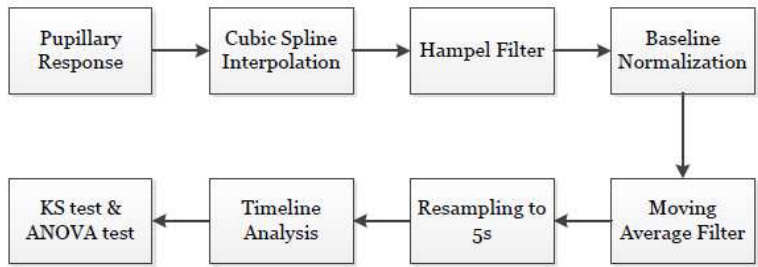

*Figure 1: Data processing steps.*

## 3   Results and Discussion

The timeline analysis of the pupillary responses is shown in Figures 2 and 3. It can be seen that although they started from same point, they follow different paths during the 5 seconds analysis window. In the case of paired video samples, the difference between real and posed smiles started from about 1.5s and continues until the end of our analysis window, (i.e. 5s) and the difference is gradually higher with the time as shown in Figure 2 (left). The KS test shows that there are significant differences between real and posed smiles stimuli ($p<0.001$).

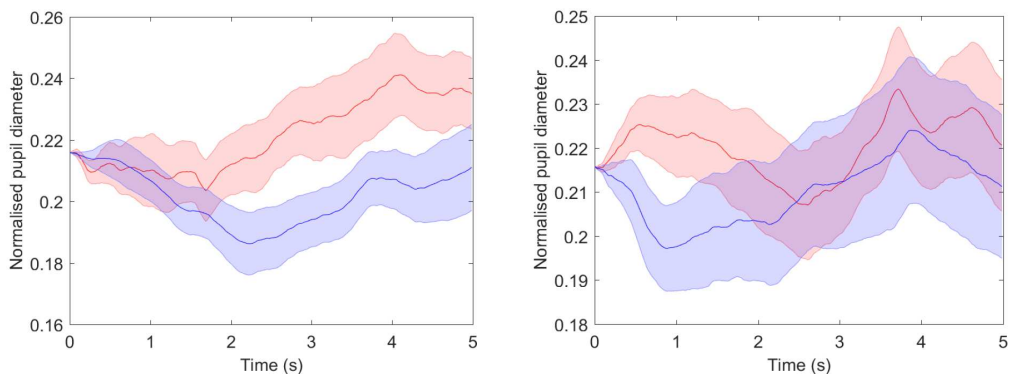

*Figure 2: Average pupil diameter timelines for paired videos (left) and paired images (right) - Real (red) vs Posed (blue) smiles.*

For paired images, big differences between real and posed smiles were seen from 0.2s to 2.2s. It can also be seen that the difference is also found after 3.3s, but not as high as during first 2s as seen in Figure 2 (right). It is due to the differences between videos and images. Overall, the KS test shows that there is significant difference between real and posed smiles for this case (p<0.001). For single (image and video) conditions, it has been shown that there is no big difference between real and posed smiles such as seen in the paired conditions as shown in Figure 3. In the case of single video, there are 7 crossing points between real and posed smiles during the 5s timeline analysis (please see Figure 3 (left)). The 7th crossing point is found at 3.8s. After that, the difference between real and posed smiles increases until the end of our analysis window. The KS test shows that there is a significant difference between real and posed smiles for single video stimuli (p<0.05). In case of single image conditions, it is very hard to differentiate between real and posed smiles by naked eye as shown in Figure 3 (right). In SV and SI, there are few overlaps on the pupillary responses as observers found it difficult to decide whether the observed smiles are real or posed. The timeline analysis shows that the pupillary responses for both smiles follow similar trends with little difference, although KS-test shows a significant difference for SI (p<0.05).

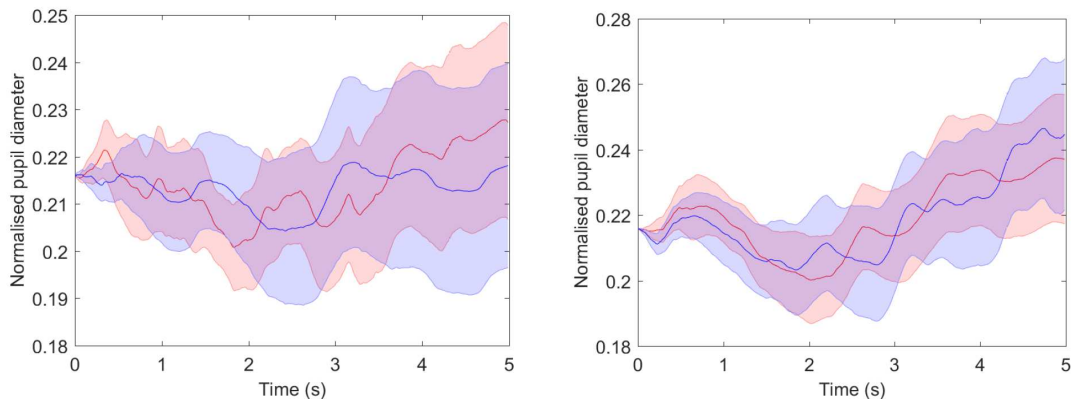

*Figure 3: Average pupil diameter timelines for single videos (left) and single images (right) - Real (red) vs Posed (blue) smiles.*

As an additional analysis, the ANOVA test shows that there is a significant difference between real and posed smiles only for PV condition (p = 0.04) where for other conditions they are not significant (p = 0.45, 0.95, and 0.97 were found for PI, SV, and SI conditions respectively). Thus, we recommend using PV condition while differentiating any emotion / facial expression from observers' pupillary responses. Further, we check the differences on mean values considering all 4 conditions as shown in Figure 4. It can be seen from Figure 4 that there is a big difference considering paired condition compared to single condition. In a nutshell, any paired condition is better than any single condition for differentiating between real and posed smiles from observers' pupillary responses.

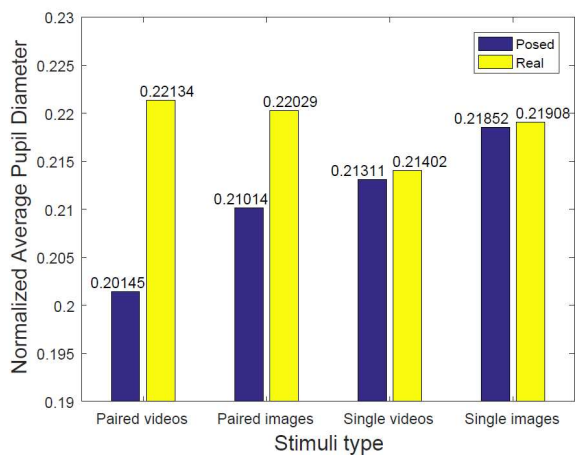

*Figure 5: Average pupil diameters for all conditions*

## 3.1 Conclusion

We ran experiments to differentiate between real and posed smiles from observers' pupillary responses. Several pre-processing steps were completed to remove unwanted noise from recorded signals. To identify the most suitable condition from the defined conditions (namely paired video, paired image, single video, and single image), we conducted timeline analyses. Although KS test and timeline analyses show that it is possible to differentiate considering each condition separately, the ANOVA test shows significant difference for paired videos only. From the result we can easily find that peoples' reactions to 'paired' conditions are more obviously different than 'single' conditions. The literature shows that video stimuli can arouse stronger emotional states than images (Horvat et al. 2015). Our analysis also supports that the paired video condition shows more significant differences than other conditions.

Our findings in this study can be potentially applicable in many real-world situations, such as using mobile applications for monitoring a caller's smiles by the recipient to enhance their emotional communication. Our future research direction will include more datasets incorporating other physiological responses (such as galvanic skin response, blood volume pulse, heart rate etc.). We will include participants from a more diverge age range and different social backgrounds. We will also use videos and images from real world scenarios, for example online tutoring and advertising websites (Gregor et al. 2014).